\documentclass{aastex63}

\usepackage[caption=false]{subfig}
\usepackage{graphicx}

\usepackage{todonotes}
\usepackage[caption=false]{subfig}
\usepackage{graphicx}
\usepackage{longtable}
\usepackage{multirow}
\usepackage{graphicx}
\usepackage{booktabs}
\usepackage{amsmath,amssymb,amsthm,txfonts}  
\usepackage[T1]{fontenc}
\usepackage{threeparttable}
\usepackage{color}
\usepackage{CJKutf8}

\newcommand{\Ni}{{$^{56}$Ni }}
\newcommand{\Co}{{$^{56}$Co }}
\newcommand{\Fe}{{$^{56}$Fe }}
\newcommand{\kg}{{$\kappa_{\rm \gamma}$ }}

\begin{document}

\title{Modeling the Multi-band Light Curves of the Afterglows of Three Gamma-Ray Bursts and Their Associated Supernovae}

\correspondingauthor{Shan-Qin Wang \begin{CJK*}{UTF8}{gbsn}(王善钦)\end{CJK*}}
\email{shanqinwang@gxu.edu.cn}

\author{Ji-Shun Lian \begin{CJK*}{UTF8}{gbsn}(连纪顺)\end{CJK*}}
\affiliation{Guangxi Key Laboratory for Relativistic Astrophysics,
School of Physical Science and Technology, Guangxi University, Nanning 530004,
China}

\author{Shan-Qin Wang\begin{CJK*}{UTF8}{gbsn}(王善钦)\end{CJK*}}
\affiliation{Guangxi Key Laboratory for Relativistic Astrophysics,
School of Physical Science and Technology, Guangxi University, Nanning 530004,
China}

\author{Wen-Pei Gan \begin{CJK*}{UTF8}{gbsn}(甘文沛)\end{CJK*}}
\affiliation{Guangxi Key Laboratory for Relativistic Astrophysics,
School of Physical Science and Technology, Guangxi University, Nanning 530004,
China}

\author{Jing-Yao Li \begin{CJK*}{UTF8}{gbsn}(李京谣)\end{CJK*}}
\affiliation{Guangxi Key Laboratory for Relativistic Astrophysics,
School of Physical Science and Technology, Guangxi University, Nanning 530004,
China}

\author{En-Wei Liang \begin{CJK*}{UTF8}{gbsn}(梁恩维)\end{CJK*}}
\affiliation{Guangxi Key Laboratory for Relativistic Astrophysics,
School of Physical Science and Technology, Guangxi University, Nanning 530004,
China}

\begin{abstract}

There are some dozen supernovae (SNe) associated with long Gamma-ray bursts (GRBs) have
been confirmed. Most of previous studies derive the physical properties
of the GRB-SNe by fitting the constructed (psuedo-)bolometric light curves.
However, many GRB-SNe have only a few filter data, for which the (psuedo-)bolometric
light curves are very difficult to be constructed. Additionally, constructing
(psuedo-)bolometric light curves rely on some assumptions.
In this paper, we use the multi-band broken power-law plus $^{56}$Ni
model to fit the multi-band light curves of the afterglows and the SNe
(SN~2001ke, SN~2013dx, and SN~2016jca) associated with three GRBs (GRB~011121, GRB~130702A,
and GRB~161219B).
We find our model can account for the multi-band light curves of the three GRB-SNe
(except for the late-time $z-$band light curves of two events), indicating that the model is
a reliable model.
The $^{56}$Ni masses we derive are higher than that in the literature.
This might be due to the fact that the $^{56}$Ni masses in the literature are usually obtained
by fitting the psuedo-bolometric light curves whose luminosities are usually
(significantly) underestimated.
We suggest that the multi-band model can not only be used to fit the multi-band light
curves of GRB-SNe that have many filter observations, but also fit those having sparse data.

\end{abstract}


\section{Introduction}
\label{sec:intro}

Gamma-ray bursts (GRBs) are the most powerful explosions in the universe.
It is widely believed that GRBs come from the relativistic jet launched by the
central engine \citep{Woosley2011}. The interactions between the jets with the
surrounding medium would produce X-ray, {UV-optical-NIR} and radio afterglows
(see \citealt{Zhang2018} and references therein).
According to the observation of prompt emission duration, GRBs is divided into long-duration bursts (LGRBs)
and short-duration bursts (SGRBs) with a dividing line of $\sim$2 seconds \citep{Kouveliotou1993}.
The observations and analysis for some dozen supernovae (SNe) associated with LGRBs
\citep{Hjorth2003,Matheson2003,Stanek2003,Malesani2004,Deng2005,Modjaz2006,Mirabal2006,Sollerman2006,
Campana2006Natur,Maeda2007,Chornock2010,Starling2011,Olivares2012,Bufano2012,Melandri2012,Singer2013,
Schulze2014,Melandri2014GRB130427A,Toy2016,D-Elia2015,Cano2017GRB161219B,Volnova2017,
Ashall2019,Melandri2019,Hu2021} indicate that most LGRBs are produced by the explosions of massive stars.
On the other hand, the confirmation of SSS17a/AT2017gfo which is a kilonova associated with GW170817 that
is a gravitational wave emitted by a merger of a neutron star binary and GRB~170817A that is an SGRB
\citep{Arcavi2017Natur,Shappee2017,Abbott2017PhRvL,Coulter2017,
Shappee2017} supports the conjecture that at least a fraction of SGRBs are produced by the
mergers of compact binary stars.

The SNe associated with LGRBs are called GRB-SNe \citep{Woosley2006,Hjorth2012,Cano2017}.
On average, one or two GRB-SNe can be found every year.
To date, there are about 60 LGRBs that have been confirmed to be associated with SNe.
Almost all GRB-SNe are broad-lined Ic (Ic-BL) SNe whose optical spectra are hydrogen-deficient
and show broad absorption line features. The spectral features indicate that the progenitors of
GRB-SNe are highly stripped, and might be Wolf-Rayet stars \citep{Price2002,Sonbas2008}.
The broad absorption lines are indicative of huge ejecta velocities $\gtrsim 2\times10^{9}$ cm s$^{-1}$.
Therefore, a major fraction of GRB-SNe (and the SNe Ic-BL without accompanying GRBs)
become so-called ``hypernovae$"$ (HNe) whose kinetic energy is $\gtrsim 10^{52}$ erg,
which is about 10 times that of normal SNe. The explosion mechanisms of GRB-SNe are still elusive.

The most prevailing model adopted to account for the light curves of GRB-SNe is
\Ni cascade decay (\Ni$\rightarrow$\Co$\rightarrow$\Fe) model (the \Ni model,
\citealt{Arnett1979,Arnett1980,Arnett1982,Arnett1996}).
Some very luminous GRB-SNe cannot be explained by the \Ni model, and
alternative or additional energy sources (e.g., the magnetar spinning-down, the
fall-back accretion, etc.) are employed to account for the light curves.

Previous studies focusing on GRB-SNe usually construct the psuedo-bolometric light
curves of the SNe and derive the physical properties of GRB-SNe by fitting the
constructed psuedo-bolometric light curves. It should be noted that, however,
the process constructing the psuedo-bolometric light curves might underestimate
the luminosities of the SNe and therefore underestimate the \Ni masses.

Recently, the model directly fit the multi-band light curves \citep{Nicholl2017}
have been adopted to fit the light curves of superluminous SNe
\citep{Nicholl2017,Moriya2018}, the tidal disruption events \citep{Mockler2019},
the luminous rapidly evolving optical transients \citep{Wang2019},
and ordinary SNe Ib and Ic \citep{Wang2022}.

In this paper, we collect published data of the {(UV-)}optical--NIR counterparts
(GRB~011121/SN~2001ke, GRB~130702A/SN~2013dx, GRB~161219B/SN~2016jca)
of {three} GRBs (GRB~011121, GRB~130702A, GRB~161219B; here, the GRBs represent their
afterglows) and use the broken power-law plus \Ni model to fit their multi-band
light curves. In Section \ref{Sec:fits}, we model the multi-band light curves of the {three} GRB-SNe
using the \Ni model. In Section \ref{Sec:discuss}, we
compare parameters and the bolometric properties of the SNe to that in the literature.
We draw some conclusions in Section \ref{Sec:Conclusions}.
The values of of the foreground {reddening} of the Milky Way ($E(B-V)_{\rm MW}$) are from \cite{Schlafly2011}.
{The standard cosmological parameters}
{($\Omega_{m}$ = 0.315, $\Omega_{\lambda}$ = 0.685, and $H_{0}$ = 67.3 $\rm kms^{-1}Mpc^{-1}$, \citealt{PlanckCollaboration2014})}
{are adopted throughout this paper.}

\section{Modeling the Multi-band Light Curves of {Three} GRB-SNe Using the \Ni Model}
\label{Sec:fits}

The information of the {GRB~011121/SN~2001ke, GRB~130702A/SN~2013dx, GRB~161219B/SN~2016jca}
are list in Table 1.
{The flux of the host galaxy of GRB~011121/SN~2001ke is negligible, and the flux of the host galaxies
of GRB~130702A/SN~2013dx and GRB~161219B/SN~2016jca have been subtracted (the magnitudes of the two
host galaxies are from \cite{Volnova2017} and \cite{Laskar2018}, respectively, and listed in Table \ref{tab:host}).}
{Then the flux of the (UV-)optical-NIR counterpart of a GRB-SN ($F_{\nu,\rm tot}(t)$) can be divided into that of the GRB
afterglow (${F}_{\nu,\rm AG}(t)$) and that of the SN (${F}_{\nu,\rm SN}(t)$) associated with the GRB, i.e.,
$F_{\nu,\rm tot}(t)={F}_{\nu,\rm AG}(t)+{F}_{\nu,\rm SN}(t)$}.

The {flux density of an} afterglow is proportional to a broken power-law decay function
($F_{\nu,\rm AG}(t)\propto(({t}/{t_{\rm {b}}})^{\alpha_{1} \cdot n}+({t}/{t_{\rm {b}}})^{\alpha_{2} \cdot n})^{{-1}/{n}}$,
{\citealt{Beuermann1999A&A}, in the representation of \citealt{Zeh2004})
and ${\nu}^{-\beta}$ ($F_{\nu,\rm AG}(t)\propto{\nu}^{-\beta}$), and can be expressed as}
$F_{\nu,\rm AG}(t)={A_{\rm AG}} \cdot(({t}/{t_{\rm {b}}})^{\alpha_{1} \cdot n}+({t}/{t_{\rm {b}}})^{\alpha_{2} \cdot n})^{{-1}/{n}} \cdot {\nu}^{-\beta}$.
{The definitions of $\alpha_1$, $\alpha_2$, $t_{\rm b}$, $n$, and $\beta$ are presented in
Table \ref{tab:parameters}.}

{We assume that the SN associated with a GRB was powered by \Ni cascade decay.
The bolometric luminosity of \Ni-powered SNe powered is (see e.g., \citealt{Arnett1982,Cha2012,Wang2015,Wanggan2022})
\begin{eqnarray}
L_{\rm SN}(t)=\frac{2}{\tau_{m}}e^{-\frac{t^{2}}{\tau_{m}^{2}}}\int_0^t e^{\frac{t'^{2}}{\tau_{m}^{2}}}
\frac{t'}{\tau_{m}}\left(\epsilon_{\rm Ni}M_{\rm Ni}e^{-{t}/{\tau_{\rm Ni}}} + \epsilon_{\rm Co}M_{\rm Ni}\frac{e^{-{t}/{\tau_{\rm Co}}}-e^{-{t}/{\tau_{\rm Ni}}}}{1-{\tau_{\rm Ni}}/{\tau_{\rm Co}}}\right)\left(1-e^{-\tau_{\gamma}(t)}\right)dt',
\label{equ:lum}
\end{eqnarray}
here, $\tau_{m}=({2\kappa M_{\rm ej}}/{\beta_{\rm SN}{v_{\rm ph}}c})^{1/2}$ is the diffusion timescale,
$\tau_{\gamma}(t)={3\kappa_{\gamma}M_{\rm ej}}/{4\pi {v_{\rm ph}}^2t^2}$ is the optical depth to $\gamma$-rays \citep{Cha2009,Cha2012}.
$\kappa$ is the optical opacity of the ejecta which is set to be 0.07 cm$^2$ g$^{-1}$,
$c$ is the speed of light, $\beta \simeq 13.8$ is a constant \citep{Arnett1982},
$\epsilon_{\rm Ni}~=~3.9\times 10^{10}$~erg~s$^{-1}$~g$^{-1}$ \citep{Sut1984,Cap1997},
$\tau_{\rm Ni}~=~$8.8~days, $\epsilon_{\rm Co}=6.8 \times 10^{9}$ erg~s$^{-1}$~g$^{-1}$
\citep{Mae2003}, $\tau_{\rm Co}~=~$111.3~days.}

{Assuming the early-time photosphere radii of the SNe is proportional to the
time, and the ejecta cool to constant temperatures ($T_{\rm f}$),
the temperatures and radii can be given by \citep{Nicholl2017}:
\begin{eqnarray}
T_{\rm ph}(t) =
\left\{
\begin{array}{lr}
\left(\frac{L_{\rm SN}(t)}{4 \pi \sigma v_{\rm ph}^2 t^2}\right)^{\frac{1}{4}},\ &
\quad \left(\frac{L_{\rm SN}(t)}{4 \pi \sigma v_{\rm ph}^2 t^2}\right)^{\frac{1}{4}} > T_{\rm f} \\
T_{\rm f},&
\left(\frac{L_{\rm SN}(t)}{4 \pi \sigma v_{\rm ph}^2 t^2}\right)^{\frac{1}{4}} \le T_{\rm f} \\
\end{array}
\right.
\end{eqnarray}
\begin{eqnarray}
R_{\rm ph}(t) =
\left\{
\begin{array}{lr}
v_{\rm ph} t, &
\quad \left(\frac{L_{\rm SN}(t)}{4 \pi \sigma v_{\rm ph}^2 t^2}\right)^{\frac{1}{4}} > T_{\rm f} \\
\left(\frac{L_{\rm SN}(t)}{4 \pi \sigma T_{\rm f}^4}\right)^{\frac{1}{2}},&
\left(\frac{L_{\rm SN}(t)}{4 \pi \sigma v_{\rm ph}^2 t^2}\right)^{\frac{1}{4}} \le T_{\rm f} \\
\end{array}
\right.
\end{eqnarray}}

To fit the multi-band light curves of the SN components, we suppose that the spectral energy distributions (SEDs) of the SNe
can be described by the {UV absorbed} blackbody model {\citep{Pra2017,Nicholl2017}},
{\begin{eqnarray}
F_{\nu,{\rm SN}} =
\left\{
\begin{array}{lr}
\big(\frac{\lambda}{\lambda_{\rm CF}}\big)(2 {\pi} h{\nu}^3/c^2)(e^{\frac{h{\nu}}{k_{\rm b}T_{\rm ph}}}-1)^{-1}\frac{R_{\rm ph}^2}{D_L^2},
\ & \quad \lambda \leq \lambda_{\rm CF} \\
(2 {\pi} h{\nu}^3/c^2)(e^{\frac{h{\nu}}{k_{\rm b}T_{\rm ph}}}-1)^{-1}\frac{R_{\rm ph}^2}{D_L^2},
\ & \quad \lambda > \lambda_{\rm CF} \\
\end{array}
\right.
\end{eqnarray}
here, $\lambda_{\rm CF}=3000$ {\AA} is the cutoff wavelength \citep{Pra2017,Nicholl2017}.}

The definitions, the units, and the priors of the parameters of the model
are listed in Table \ref{tab:parameters}. {The values of $A_{\rm V,host}$ (or $E(B-V)_{\rm host}$)
of GRB~130702A/SN~2013dx and GRB~161219B/SN~2016jca have been given by the literature, and can be set
to be constants. Hence, the multi-band \Ni model fitting GRB~130702A/SN~2013dx and GRB~161219B/SN~2016jca
has five free parameters ($M_{\rm ej}$, $v_{\rm ph}$, $M_{\rm Ni}$, $\kappa_{\gamma}$, and $T_{\rm f}$).
GRB~011121/SN~2001ke is far away from the host galaxy \citep{Bloom2002GRB011121}; \cite{Greiner2003}
suggest that it has no host galaxy extinction, while \cite{Yolda2007} get an upper limit of $E(B-V)_{\rm host}$
(0.08 mag). We assume that $A_{\rm V,host}$ of GRB~011121/SN~2001ke is an additional free parameter whose range
is 0 to 0.248 mag. Additionally, we assume that the ratio of $M_{\rm Ni}$ to $M_{\rm ej}$ is $\leq0.2$ \citep{Ume2008}.}
We adopt the Markov Chain Monte Carlo (MCMC) method by using \texttt{emcee} of Python package \citep{Foreman-Mackey2013}
to fit the data to obtain the best-fitting parameters and 1 $\sigma$ parameter range.

\begin{figure}[tbph]
\begin{center}
\includegraphics[angle=0,width=0.45\textwidth]{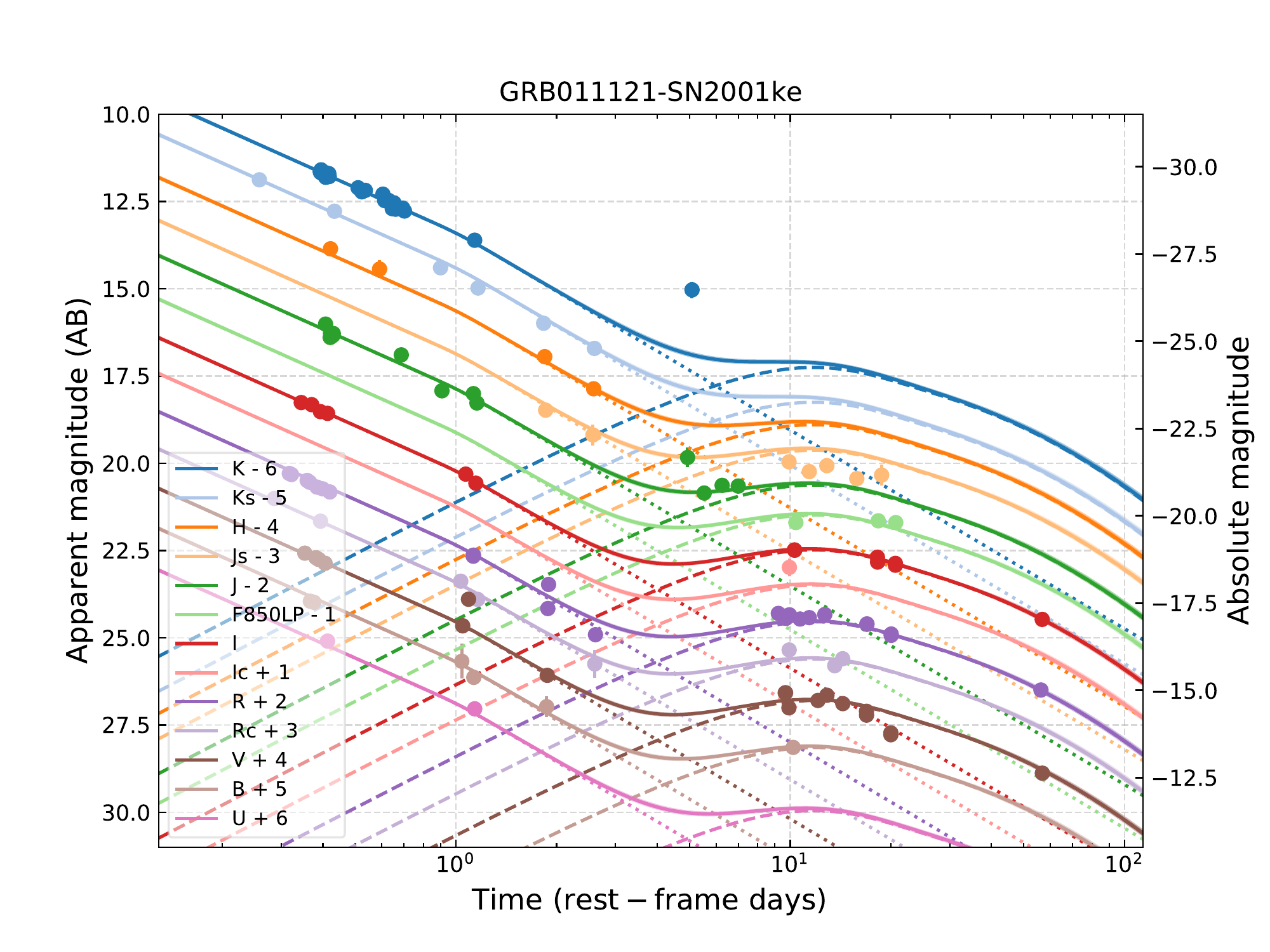}
\includegraphics[angle=0,width=0.45\textwidth]{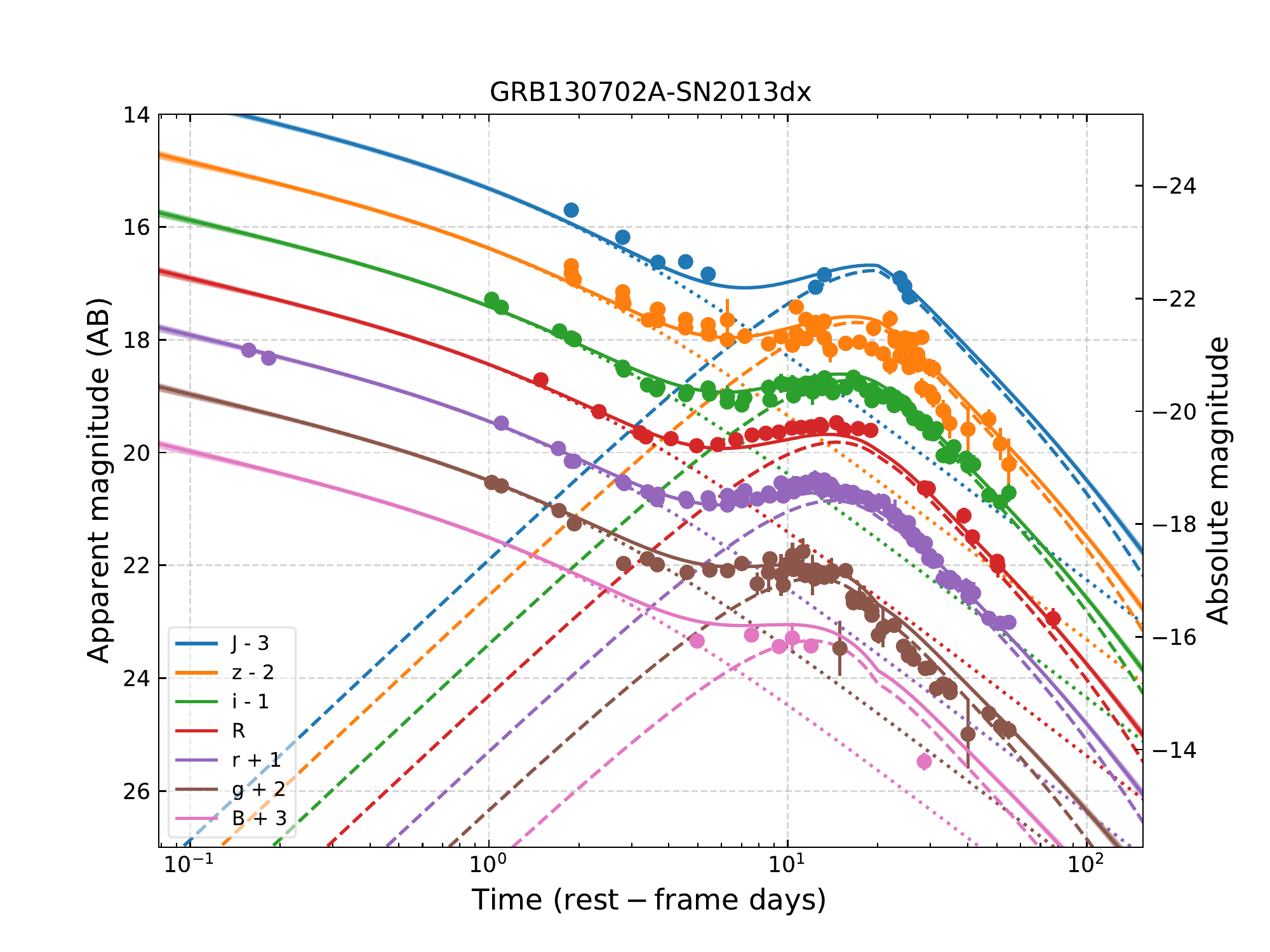}
\includegraphics[angle=0,width=0.45\textwidth]{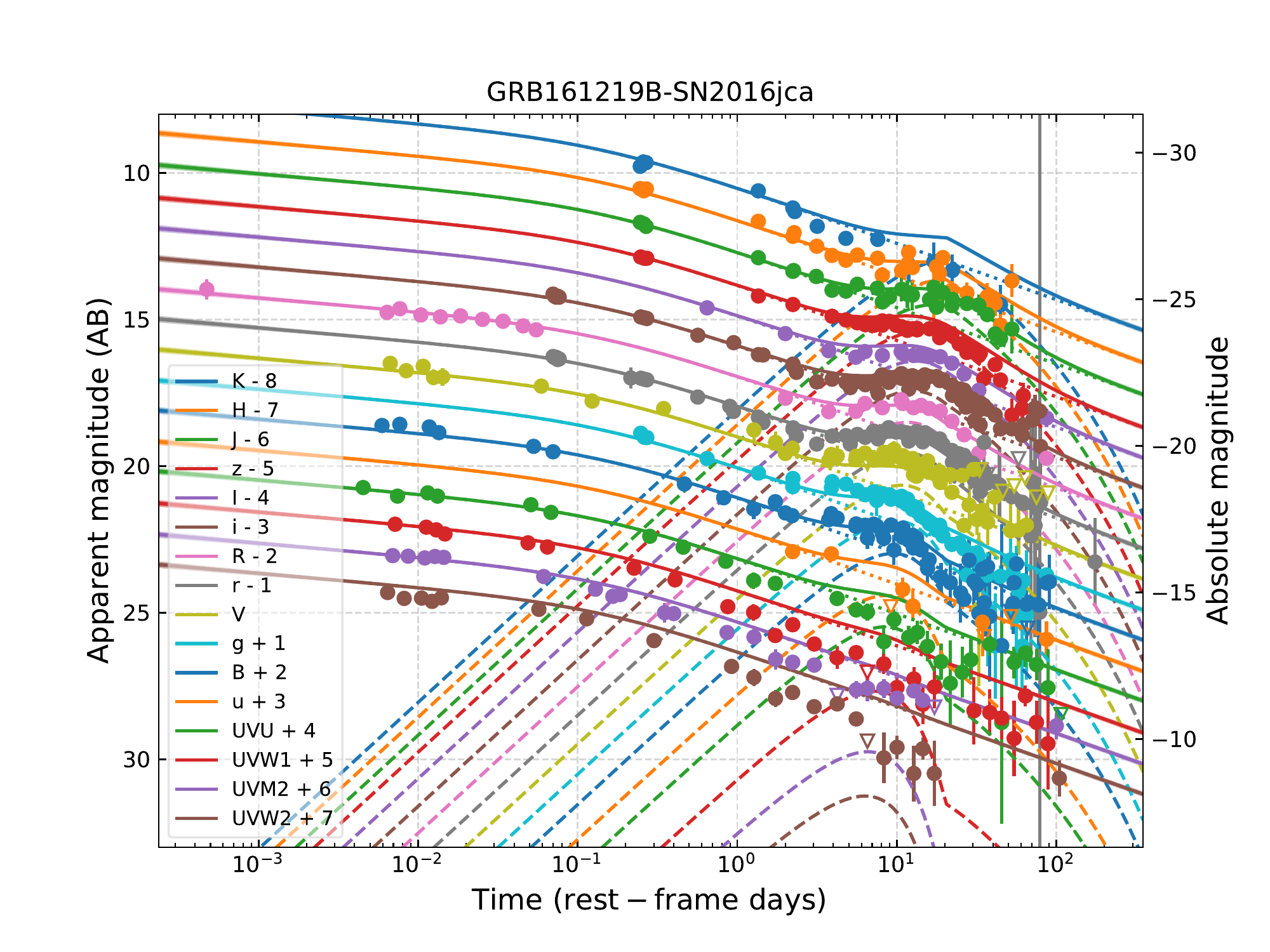}
\includegraphics[angle=0,width=0.45\textwidth]{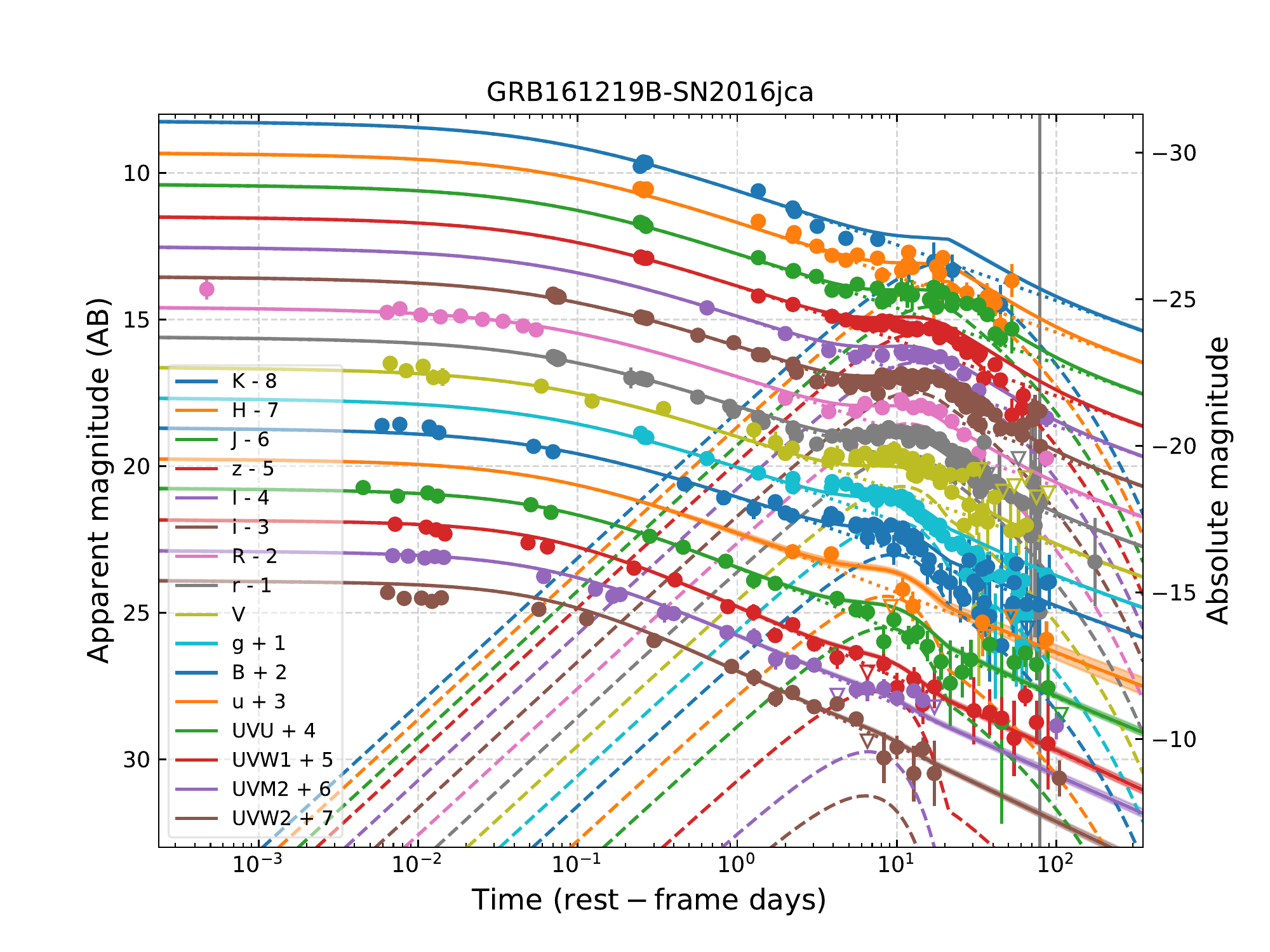}
\end{center}
\caption{{The fits of the multi-band light curves of GRB~011121/SN~2001ke (the top-left panel),
GRB~130702A/SN~2013dx (the top-right panel)
and GRB~161219B--SN~2016jca (the bottom panels). The fit represented in the bottom-left panel
based on the assumption that the values of $\alpha_2$ in all bands are the same one. In contrast,
the fit represented in the bottom-right panel assumes that the values of $\alpha_2$ in
optical-NIR bands and UV bands are different, and the $\alpha_2$ of UV bands are different from each other.
The solid, dotted and the dashed lines represent the total flux, the afterglow flux, and the SN flux, respectively.}}
\label{fig:fits}
\end{figure}

The fits of the {three} GRB-SNe and the best-fit parameters are presented in Figure \ref{fig:fits} and Table \ref{tab:para-best}, respectively.
The corresponding corner plots are shown in {Figures \ref{corner_GRB011121}, \ref{corner_GRB130702A}, and \ref{corner_GRB161219B}} in the Appendix.
{All or most optical and NIR} bands of the {three} GRB-SNe can
be fitted by the multi-band model, except for
the {late-time} $z-$band light curve of GRB~130702A/SN~2013dx and
GRB~161219B/SN~2016jca which cannot be well fitted by the multi-band model (see
{the top and bottom-left panels of} Figure \ref{fig:fits}).

{The UV band light curves cannot be well fitted by the model.
As shown in Figure 2 of \cite{Laskar2018}, the values of $\alpha_{2}$ of different UV
bands of GRB~161219B/SN~2016jca are different. Hence, assuming a same value
of $\alpha_{2}$ for all bands can result in a bad fit. To improve the fit, we assume
that the values of $\alpha_{2}$ in different UV bands are different from each other,
and different from the value in optical and NIR bands. The new fit for the light curves of GRB~161219B/SN~2016jca
and the corresponding corner plot are presented in the bottom-right panel of Figure \ref{fig:fits}
and Figure \ref{corner_GRB161219B-new}, respectively. The parameters of the new fit are listed
in the last column of Table \ref{tab:para-best}. We find that the new fit is
better than the first fit since the UV bands are also well matched by the model.}

There are {two} (possible) reasons that can explain the bad quality of the fits for
{$z-$band light curves of the two GRB-SNe}.
(1). {Their late-time} $z-$band light curves show fluctuation features
that cannot be fully fitted by the theoretical light curves which are smooth.
{(2). Their late-time SEDs deviate the blackbody function in $z-$band}.

The {derived} masses of \Ni of SN~2001ke, SN~2013dx and SN~2016jca are ${0.46}\pm0.01$ M$_\odot$,
${0.74}\pm0.01$ M$_\odot$, and ${0.33}\pm0.00$ M$_\odot$, respectively.
The ejecta masses of the {three}
GRB-SNe are ${4.02}^{+0.53}_{-0.58}$ M$_\odot$, ${3.71}\pm0.03$ M$_\odot$ and ${1.64}\pm0.02$  M$_\odot$,
respectively. The respective velocity of the ejecta of the {three} GRB-SNe are ${4.22}^{+0.54}_{-0.61}\times10^9$ cm s$^{-1}$,
${2.61}\pm0.02\times10^9$ cm s$^{-1}$, and ${2.17}\pm0.03\times10^9$ cm s$^{-1}$.
The parameters are roughly consistent with the parameter ranges in the literature.

\section{Discussion}
\label{Sec:discuss}

Here, we compare the values of the \Ni masses, the ejecta masses, the ejecta velocity,
and the kinetic energy of the ejecta of the {three} GRB-SNe to that
in the literature and discuss the reasons causing the discrepancies.
Moreover, we discuss the theoretical bolometric light curves of
the {three} GRB-SNe.

\subsection{The \Ni Masses of the {Three} GRB-SNe}

The \Ni mass of GRB~011121/SN~2001ke is ${0.46}\pm0.01$ M$_\odot$, we do not find the literature's value.
The \Ni mass of GRB~130702A/SN~2013dx is ${0.74}\pm0.01$ M$_\odot$, which is
$\sim{2.0}$ and $\sim{3.7}$ times {those} of the values derived by \cite{Toy2016} ($0.37\pm0.01$ M$_\odot$)
and \cite{D-Elia2015} ($0.2$ M$_\odot$).
The \Ni mass of GRB~161219B/SN~2016jca is ${0.33}\pm0.00$ M$_\odot$,
which is $\sim {1.50}_{-0.40}^{+0.86}$ and $\sim {1.22}_{-0.19}^{+0.28}$ times {those}
of the values derived by \cite{Cano2017GRB161219B} ($0.22\pm0.08$ M$_\odot$) and \cite{Ashall2019} ($0.27\pm0.05$ M$_\odot$),
respectively.
The discrepancy might be due to the facts that \cite{Toy2016}, \cite{Cano2017GRB161219B} and
\cite{Ashall2019} derived the \Ni masses by fitting the psuedo-bolometric light curves
\footnote{\cite{D-Elia2015} construct the psuedo-bolometric light curve of SN~2013dx and
derive the \Ni mass by scaling the psuedo-bolometric light curve of SN~2003dh.}
which are dimmer than the bolometric light curves
and that our blackbody multi-band fit{s correspond} to the bolometric light curve{s}.

The \Ni {mass} of GRB~130702A/SN~2013dx {is} rather large, but
comparable to the \Ni mass of SN~1998bw which is $0.4-0.7$ M$_\odot$ \citep{Iwamoto1998Natur,Nakamura2001} or
$0.54^{+0.08}_{-0.07}$ M$_\odot$ \citep{Lyman2016}. Therefore,
we suggest that the \Ni mass {is} reasonable.

\subsection{The Properties of the Ejecta}


The ejecta mass of SN~2013dx and SN~2016jca are
${3.71}\pm0.03$ M$_\odot$ and ${1.64}\pm0.02$ M$_\odot$,
which are lower than the values derive by the literature ($3.1\pm0.1$ M$_\odot$ \citep{Toy2016} or $\sim7$ M$_\odot$ \citep{D-Elia2015} for
SN~2013dx, $5.8\pm0.3$ M$_\odot$ \citep{Cano2017GRB161219B} or $6.5 \pm 1.5$ M$_\odot$ \citep{Ashall2019} for SN~2016jca).


{For example,
\cite{Toy2016} and \cite{D-Elia2015} assume that the velocity of SN~2013dx are $2.13\times10^9$ cm s$^{-1}$
and $\sim 2.9\times10^9$ cm s$^{-1}$ \footnote{This value is derived from the medians of the ejecta mass (7 M$_\odot$) and kinetic energy
($35\times10^{51}$ erg) provided by \cite{D-Elia2015}, assuming $E_{\rm k}=\frac{3}{10} M_{\rm ej} v_{\rm ph}^{2}$.}, respectively;
\cite{Cano2017GRB161219B} and \cite{Ashall2019} assume that the velocity of SN~2016jca
are $2.97\pm0.15\times10^9$ cm s$^{-1}$ and $3.5\pm0.7\times10^9$ cm s$^{-1}$, respectively.}

{Our derived early-time photospheric} velocities of SN~2013dx and SN~2016jca are ${2.61}\pm0.02\times10^9$ cm s$^{-1}$
and ${2.17}\pm0.03\times10^9$ cm s$^{-1}$. {The former} is
{between the two values adopted by \cite{Toy2016} (${2.13}\times10^9$ cm s$^{-1}$)}
and \cite{D-Elia2015} ($2.7\times10^9$ cm s$^{-1}$); {the
latter is} lower than those derived by \cite{Cano2017GRB161219B} ($2.97\pm0.15\times10^9$ cm s$^{-1}$),
and \cite{Ashall2019} ($3.5\pm0.7\times10^9$ cm s$^{-1}$).
\footnote{The SN velocities inferred from the spectra evolve (usually decrease) with the time.
\cite{Toy2016} find that the spectral velocity of SN~2013dx inferred from the
SiII lines at days 9.3, 11.3, 14.2, 31.3, 33.3 are 2.81, 2.52,
2.13, 1.17, and 1.08 $\times10^9$ cm s$^{-1}$, respectively. \cite{D-Elia2015} find that the velocity
of SN~2013dx decline from $\sim$2.7 $\times10^9$ cm s$^{-1}$ at day 8 to $\sim$3.5 $\times10^8$ cm s$^{-1}$ at day 40.
Previous studies fitting the (psuedo-)bolometric light curves usually adopt the velocity derived from the
spectra obtained around maximum light or earlier epochs.}


The kinetic energy ($E_{\rm k}=\frac{3}{10} M_{\rm ej} v_{\rm ph}^{2}$) of the ejecta of SN~2001ke
is ${4.27}^{+1.88}_{-1.60}\times10^{52}$ erg.
The kinetic energy of the ejecta of SN~2013dx is ${1.51}\pm0.04\times 10^{52}$ erg, which is {comparable to}
the value derive by \cite{Toy2016} ($8.2 \times 10^{51}$ erg) and significantly lower than
the value inferred by \cite{D-Elia2015} ($3.5 \times 10^{52}$ erg).
The kinetic energy of the ejecta of SN~2016jca is ${4.6}\pm0.2\times 10^{51}$ erg,
which is significantly lower than that derived by \cite{Cano2017GRB161219B} ($5.1\pm0.8\times 10^{52}$ erg).

\subsection{The Theoretical Bolometric Light Curves}

We use the derived best-fitting parameters to yield the bolometric light curves of the
{three} GRB-SNe we study, see Figure \ref{Bolometric luminosity}. We find that the peak bolometric
luminosities of SN~2001ke, SN~2013dx, and SN~2016jca are
${1.37}\times10^{43}$ erg s$^{-1}$, ${1.92}\times10^{43}$ erg s$^{-1}$ and ${1.04}\times10^{43}$ erg s$^{-1}$, respectively.

For comparison, the peak (psuedo-)bolometric luminosities of the {three} GRB-SNe derived by
the literature are $6\times 10^{42}$ erg s$^{-1}$ \citep{Cano2017}, $1 \times10^{43}$ erg s$^{-1}$ \citep{Toy2016}
and $6.3\times10^{42}$ erg s$^{-1}$ \citep{Ashall2019} or
$4.6\times10^{42}$ erg s$^{-1}$ \citep{Cano2017GRB161219B}, respectively.

By comparing our derived peak bolometric luminosities of SN~2001ke,
SN~2013dx, and SN~2016jca to their peak (psuedo-)bolometric luminosities in the literature,
we find that the former are respectively {2.28, 1.92, and 1.65 (or 2.26)}
times that the latter.

The discrepancies of the peak luminosities of bolometric light curves we derive and {those} of
the psuedo-bolometric light curves might be due to the fact that the latter omit the flux in
UV and/or IR bands. \cite{Toy2016} construct the psuedo-bolometric light curve of SN~2013dx by
integrating the flux in $g'r'i'z'yJ$ bands, more flux are neglected.
\cite{Cano2017GRB161219B} use the $griz$ band data to construct the psuedo-bolometric
light curve of SN~2016jca, the flux might {also} be underestimated.

Our derived rise time of SN~2001ke and SN~2013dx are respectively 11.8 days and
{13.7} days, which are respectively smaller than {and comparable to} the rise time of the {two}
SNe in the literature which are $\sim$17.5 days \citep{Cano2017} and
$\sim$14 days \citep{Toy2016}.
{Our derived rise time of SN~2016jca is {10.7} days, which is
slightly larger than in the literature which is $\sim$10 days \citep{Ashall2019}.}

\section{Conclusions}
\label{Sec:Conclusions}

In the past two decades, a few dozen LGRBs have been confirmed to be associated with SNe Ic,
most of which are SNe Ic-BL and HNe. While the kinetic energy of most GRB-SNe is $\gtrsim 10$
times that normal SNe Ic, their average peak luminosities are not significantly higher than
{those} of SNe Ic. Therefore, the \Ni model adopted to account for the light curves of normal SNe
Ic have also been used to explain the light curves of GRB-SNe. However, many studies exploring
the energy sources of GRB-SNe construct the psuedo-bolometric light curves and fit them.
This method might underestimate the \Ni masses needed to power the light curves of SNe.

{We collected photometric data} of {three} well-observed GRB-SNe (GRB~011121/SN~2001ke,
GRB~130702A/SN~2013dx, GRB~161219B/SN~2016jca) and use the {multi-band broken} power
law plus \Ni model to fit the multi-band light curves of the total flux which is the sum
of {those} of the afterglows of the GRBs and the SNe.
{The multi-band model we use fit the observed multi-band data, rather than the
psuedo-bolometric light curves constructed by taking some assumptions.
A larger dataset can pose more stringent constraints on the physical
parameters.}

{We find that the multi-band light curves of GRB~011121/SN~2001ke can be fitted {by the model we use};
the multi-band light curves of GRB~130702A/SN~2013dx and GRB~161219B/SN~2016jca
can be fitted by the model (except their late-time $z-$band light curves).}
This {indicates that the UV-optical-NIR SEDs of SNe associated with GRBs can be
well described by the UV-absorbed blackbody model, and that our model can account for the
multi-band light curves of the three GRB-SNe.}


{Our derived} \Ni masses of SN~2013dx and SN~2016jca are ${0.74}\pm0.01$ M$_\odot$
and ${0.33}\pm0.00$ M$_\odot$, respectively. The former is about $\sim {2.0}$ and
$\sim {3.7}$ times {those} ofthe values derived by \cite{Toy2016} and \cite{D-Elia2015},
while the latter is $\sim {1.50}_{-0.40}^{+0.86}$ and $\sim {1.22}_{-0.19}^{+0.28}$ times
{those} of the values derived by \cite{Cano2017GRB161219B} and \cite{Ashall2019}. This {might
be due to the fact}
that the constructed psuedo-bolometric light curves of SN~2013dx and SN~2016jca omit a fraction of
the total flux. {Therefore, w}e suggest that the $^{56}$Ni masses of at least a fraction of
GRB-SNe have been underestimated, and the multi-band \Ni model can make it possible to avoid underestimating
the luminosities of SNe and therefore the \Ni masses.

{The derived early-time photospheric velocities of SN~2013dx and
SN~2016jca are ${2.61}\pm0.02\times10^9$ cm s$^{-1}$ and ${2.17}\pm0.03\times10^9$ cm s$^{-1}$,
the former is between those adopted in the literature (2.13 or 2.7$\times10^9$ cm s$^{-1}$),
while the latter is lower than that in the literature (2.97 or 3.5 $\pm0.15\times10^9$ cm s$^{-1}$).
The derived kinetic energy of SN~2013dx and SN~2016jca are
${1.51}\pm0.04 \times 10^{52}$ erg and ${4.6}\pm0.2 \times 10^{51}$ erg.
While the former is (significantly) lower in the literature ($8.2 \times 10^{51}$ erg or $3.5 \times 10^{52}$ erg),
the latter is significantly lower than that in the literature ($5.1\pm0.8\times 10^{52}$ erg for SN~2016jca).}

{Our study demonstrate the validity of the multi-band afterglow plus \Ni model for
the the fits of multi-band light curves of GRB-SNe. The model can be
regarded as an independent model that do not rely on the (psuedo-)bolometric light curves constructed.
Although the GRB-SNe we fit have ample data at many bands,
we expect that the model can also be used to the multi-band light curves of GRB-SNe
observed in only one, two, or three bands at some or all epochs. For the GRB-SNe
with sparse data, the multi-band model can play a key role to determine their physical properties
by fitting their multi-band light curves, since constructing the (psuedo-)bolometric light curves is very difficult.}

\begin{acknowledgments}
{We thank the anonymous referee for helpful comments and
suggestions that have allowed us to improve this manuscript.}
This work is supported by National Natural Science Foundation of China
(grants 11963001, 11851304, 11973020 (C0035736), and U1938201),
the Bagui Scholars Program (LEW),
{the Special Funding for Guangxi Distinguished Professors (2017AD22006)},
and the Bagui Young Scholars Program (LHJ).

\end{acknowledgments}

\clearpage

\begin{table*}
\begin{center}
\footnotesize
\centering
{\caption{The information of GRB~011121/SN~2001ke, GRB~130702A/SN~2013dx, and GRB~161219B/SN~2016jca.}}
\label{tab:info}
\tabcolsep 10pt 
\begin{tabular}{cccccccc}
\toprule
                                  &        RA.               &        Dec.                             &   $z$             & $E(B-V)_{\rm MW}$$^a$             &     $E(B-V)_{\rm host}$    &  data sources$^e$   \\
\noalign{\smallskip}\hline\noalign{\smallskip}
GRB~011121/SN~2001ke              &  $ 11^{\rm h}34^{\rm m}29^{\rm s}.67 $ &  $  -76^{\circ}01'41''.6 $&  0.362            & 0.419   &   {$\leq$0.08$^b$ }     &  {1, 2, 3, 4  }   \\
GRB~130702A/SN~2013dx             &$14^{\rm h}29^{\rm m}14^{\rm s}.78$     & $+15^{\circ}46'26''.4$    &   0.145           & 0.024   &   {0.032$^c$}           &   5, 6, 7     \\
GRB~161219B/SN~2016jca            &$ 06^{\rm h}06^{\rm m}51^{\rm s}.37 $   &$-26^{\circ}47'29''.7$     &   0.1475          & 0.028   &  ${0.017}\pm0.012$$^d$  & {8, 9, 10, 11, 12, 13, 14}  \\
\bottomrule

\noalign{\smallskip}
\end{tabular}
\end{center}
$^a$ \cite{Schlafly2011}.\\
$^b$ \cite{Yolda2007}.\\
$^c$ \cite{Toy2016} (assuming that the value of the total to selective extinction ratio ($R_{V}$) is 3.1, which is the typical value of $R_{V}$ of the Milky Way, \citealt{Schultz1975}). \\
$^d$ \cite{Cano2017GRB161219B}.\\
$^e$ 1. \cite{Bloom2002GRB011121}; 2. \cite{Price2002GRB011121}; 3. \cite{Garnavich2003}; 4. \cite{Greiner2003};
5. \cite{Toy2016}; 6. \cite{Volnova2017}; 7. \cite{D-Elia2015};
8. \cite{Buckley2016}; 9. \cite{Mazaeva2016}; 10. \cite{Martin-Carrillo2016}; 11. \cite{Fujiwara2016}; 12. \cite{Cano2017GRB161219B}; 13. \cite{Ashall2019}; 14. \cite{Laskar2018}
({The UVOT white ($UVh$) band data are not included, since $UVh$ is not a narrow band data; however, the clear band data are included and labeled as $R-$band, since the two are closely approximated.}) \\
\end{table*}

\clearpage

{
\begin{table*}
\begin{center}
\tabcolsep 12pt 
\tabletypesize{\scriptsize}
\caption{The AB magnitudes of the host galaxies of GRB~130702A/SN~2013dx and GRB~161219B/SN~2016jca.}
\label{tab:host}
\begin{tabular}{ccccc}
\toprule
Band                          & GRB~130702A/SN~2013dx       & GRB~161219B/SN~2016jca \\
\noalign{\smallskip}\hline\noalign{\smallskip}
$K  $                          & -                                                 &  $20.73\pm0.11$ \\
$H$                             & -                                                 & $20.73\pm0.07$   \\
$J$                              &    $23.15\pm0.48$                   & $20.80\pm0.07$  \\
$z$                             &        $22.97\pm0.18$                         &  $20.67\pm0.04$    \\
$I $                          &      -                                         &$21.08\pm0.08$  \\
$i $                            &    $22.96\pm0.09$                             &    $20.86\pm0.03$      \\
$R$                         &      $23.17\pm0.06$                      & $21.33\pm0.11$ \\
$r$                              &   $23.05\pm0.06$                 & $21.13\pm0.05$  \\
$V$                            &     -                                       &  $21.03\pm0.06$ \\
$g$                              & $23.52\pm0.06$                  & $21.48\pm0.03$ \\
$B$                          &   $23.39\pm0.20$                      &   $21.63^{+0.01}_{-0.05}$ \\
$u$                           &     -                                         &  $22.79^{+0.31}_{-0.23}$\\
$UVU$                       &     -                                       &      $22.62\pm0.12$           \\
$UVW1$                       &     -                                     &   $22.95\pm0.12$\\
$UVM2$                       &    -                                      & $23.77^{+0.49}_{-0.29}$ \\
$UVW2$                       &      -                                     &   $23.41\pm0.23$ \\
\bottomrule
\end{tabular}
\end{center}
\end{table*}
}

\clearpage


\begin{table*}
\begin{center}
\footnotesize
\centering
\caption{The definitions, the units and the prior ranges of the parameters of model.}
\label{tab:parameters}
\tabcolsep 10pt 
\begin{tabular}{ccccccc}
\toprule
Parameters   & Definitions                                         &  Unit                     &   Prior    \\
\noalign{\smallskip}\hline\noalign{\smallskip}
$A$          & Parameters describing the intensity of afterglow    &                                 &    $[10^1, 10^{40}]$     \\
$\alpha_1$   & {The prebreak decay slope of the afterglow light curve}                      &              &    $[0.01,6]$$^a$            \\
$\alpha_2$   & {The postbreak decay slope of the afterglow light curve }            &          &    $[0.01, 6]$$^a$        \\
$n$           &   {The sharpness of the break}                   &                                &    $[1,30]$            \\
$t_{\rm b}$ & The break time                                       &  days                                  &    $[0.005,30]$$^a$     \\
$\beta$      & Power-law Spectral index                            &                                          &    $[0.01,4]$            \\
$M_{\rm ej}$ & The ejecta mass                                     & M$_\odot$                           &    $[1, 15]$            \\
$v_{\rm ph}$          & The {early-time photospheric velocity}                                 & $10^9$ cm s$^{-1}$     &    $[1.5, 5.0]$       \\
$M_{\rm Ni}$ & The \Ni mass                                        &   M$_\odot$                      &    $[0.1, 0.8]$              \\
$\kappa_{\gamma} $& Gamma-ray opacity of \Ni-cascade-decay photons &   cm$^2$g$^{-1}$         &  [$10^{-1.57}, 10^3] $   \\
$T_{\rm f}$  & The temperature floor of the photosphere            &   {10$^3$} K                       &    {$[1, 15] $}          \\
 {$A_{\rm V,host}$}&     {The extinction of host galaxy}    &      {mag}                    &  $[{0, 0.248}]$$^b$   \\
\bottomrule
\noalign{\smallskip}
\end{tabular}
\end{center}
$^a$ {Based on the fits of \cite{Greiner2003}, the ranges of $\alpha_1$, $\alpha_2$, and $t_{\rm b}$ of the afterglow of GRB~011121 are set to be {$[0.5, 2.5] $},{$[2.0, 4.0] $}, and {$[0.6, 1.8] $}, respectively.}\\
$^b$ {This is the range of $A_{\rm V,host}$ of GRB~011121/SN~2001ke, the values $A_{\rm V, host}$ of
GRB~130702A/SN~2013dx and GRB~161219B/SN~2016jca are constants.}
\end{table*}

\clearpage

{
\begin{table*}
\centering
\tabcolsep 10pt 
\tabletypesize{\scriptsize}
\caption{The Best-fitting Parameters of multi-band light curves of GRB~011121/SN~2001ke,
GRB~130702A/SN~2013dx, and GRB~161219B/SN~2016jca.}
\label{tab:para-best}
\begin{tabular}{cccccc}
\toprule
 Parameters                  & GRB~011121/SN~2001ke                       & GRB~130702A/SN~2013dx         & GRB~161219B/SN~2016jca  & GRB~161219B/SN~2016jca  \\
\noalign{\smallskip}\hline\noalign{\smallskip}
 $\rm \log{A}$                        &$12.74^{+0.13}_{-0.16}$                   & $4.67^{+0.42}_{-0.41}$            &  $7.36^{+0.11}_{-0.11}$&       $6.70^{+0.11}_{-0.11}$        \\
$\alpha_1$                            &  $1.71^{+0.02}_{-0.02}$                      &  $0.48^{+0.04}_{-0.05}$        &  $0.19^{+0.02}_{-0.02}$   &          $0.01^{+0.01}_{-0.00}$              \\
$\alpha_2$                             &$2.28^{+0.06}_{-0.05}$                       &  $1.64^{+0.07}_{-0.07}$        &  $0.77^{+0.01}_{-0.01}$ &                  $0.77^{+0.01}_{-0.01}$             \\
$\alpha_{2,u}$                          &               -                                       &                      -                         &             -                           &            $0.83^{+0.03}_{-0.03}$             \\
 $\alpha_{2,UVU}$                       &                           -                          &         -                                    &             -                            &        $0.89^{+0.02}_{-0.02}$          \\
$\alpha_{2,UVW1}$                      &                          -                         &                            -                    &               -                         &             $0.99^{+0.02}_{-0.02}$                  \\
$\alpha_{2,UVW2}$                    &                            -                         &                     -                           &              -                             &             $1.03^{+0.02}_{-0.02}$             \\
$\alpha_{2,UVM2}$                     &                                    -                &                             -                      &                   -                        &         $0.96^{+0.02}_{-0.02}$              \\
$n$                                             & $23.49^{+4.62}_{-6.69}$                    &$1.03^{+0.05}_{-0.02}$    & $1.96^{+0.32}_{-0.27}$        &              $1.17^{+0.06}_{-0.05}$              \\
$t_{\rm b}$ (days)                   &  $1.05^{+0.07}_{-0.06}$                      &  $2.06^{+0.44}_{-0.40}$        &  $0.15^{+0.01}_{-0.01}$  &          $0.07^{+0.00}_{-0.00}$     \\
$\beta$                                         &  $0.77^{+0.01}_{-0.01}$                   &  $0.17^{+0.03}_{-0.03}$         &  $0.32^{+0.01}_{-0.01}$&          $0.26^{+0.01}_{-0.01}$            \\
$M_{\rm ej}$ (M$_{\odot}$)           & $4.02^{+0.53}_{-0.58}$                   & $3.71^{+0.03}_{-0.03}$        &  $1.70^{+0.02}_{-0.02}$ &          $1.64^{+0.02}_{-0.02}$         \\
$v_{\rm ph}$ ($10^9$ cm s$^{-1}$)    & $4.22^{+0.54}_{-0.61}$                   &$2.61^{+0.02}_{-0.02}$      &  $2.26^{+0.03}_{-0.03}$&      $2.17^{+0.03}_{-0.03}$                    \\
$M_{\rm Ni}$ (M$_{\odot}$)           &  $0.46^{+0.01}_{-0.01}$                &    $0.74^{+0.01}_{-0.01}$        & $0.34^{+0.00}_{-0.00}$ &           $0.33^{+0.00}_{-0.00}$              \\
$\log$\kg                                   &   $-0.79^{+0.06}_{-0.07}$              & $-1.57^{+0.01}_{-0.00}$         &  $-1.25^{+0.03}_{-0.03}$  &           $-1.26^{+0.03}_{-0.03}$          \\
$T_{\rm f}$ ($10^3$K)                       & $8.03^{+0.11}_{-0.11}$                &    $5.71^{+0.04}_{-0.04}$              &$4.91^{+0.12}_{-0.09}$ &        $4.84^{+0.09}_{-0.13}$                                \\
$A_{\rm V,host}$  (mag)                    &   $0.01^{+0.01}_{-0.01}$          &                     -                                   &           -                    &                     -                                    \\
$\chi^{\rm 2}$/dof                      &       6.74                         &       13.86                                     &   13.57                                    &             12.38                                             \\
\bottomrule
\end{tabular}
\end{table*}
}

\clearpage

{
\begin{figure}[tbph]
\begin{center}
\includegraphics[angle=0,width=0.8\textwidth]{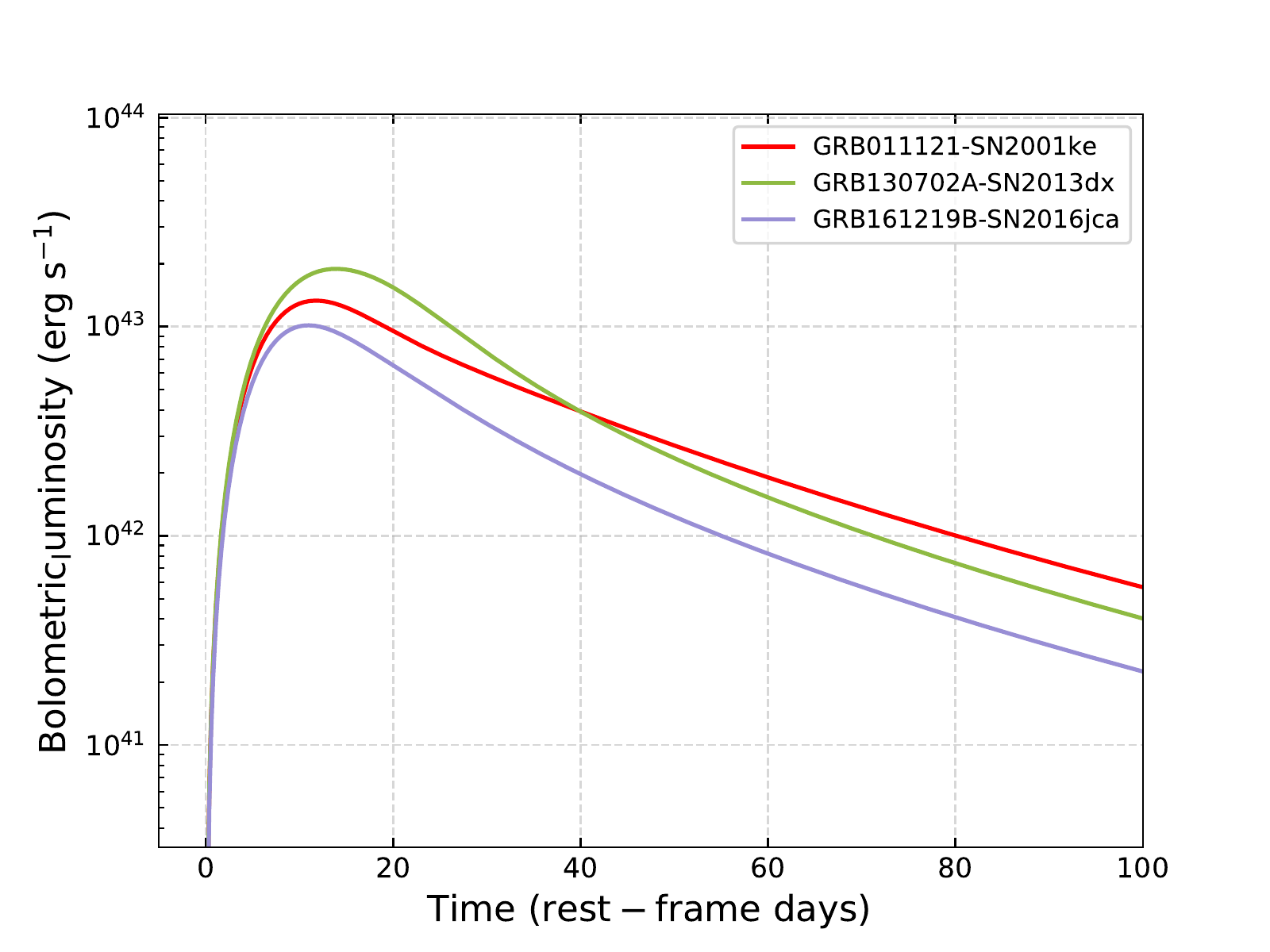}
\end{center}
\caption{The bolometric light curves reproduced by the best-fitting parameters of the \Ni model.}
\label{Bolometric luminosity}
\end{figure}
}
\clearpage

\appendix
\setcounter{table}{0}
\setcounter{figure}{0}
\setcounter{equation}{0}
\renewcommand{\thetable}{A\arabic{table}}
\renewcommand{\thefigure}{A\arabic{figure}}
\renewcommand\theequation{A.\arabic{equation}}



\begin{figure}[tbph]
\begin{center}
\includegraphics[angle=0,width=0.8\textwidth]{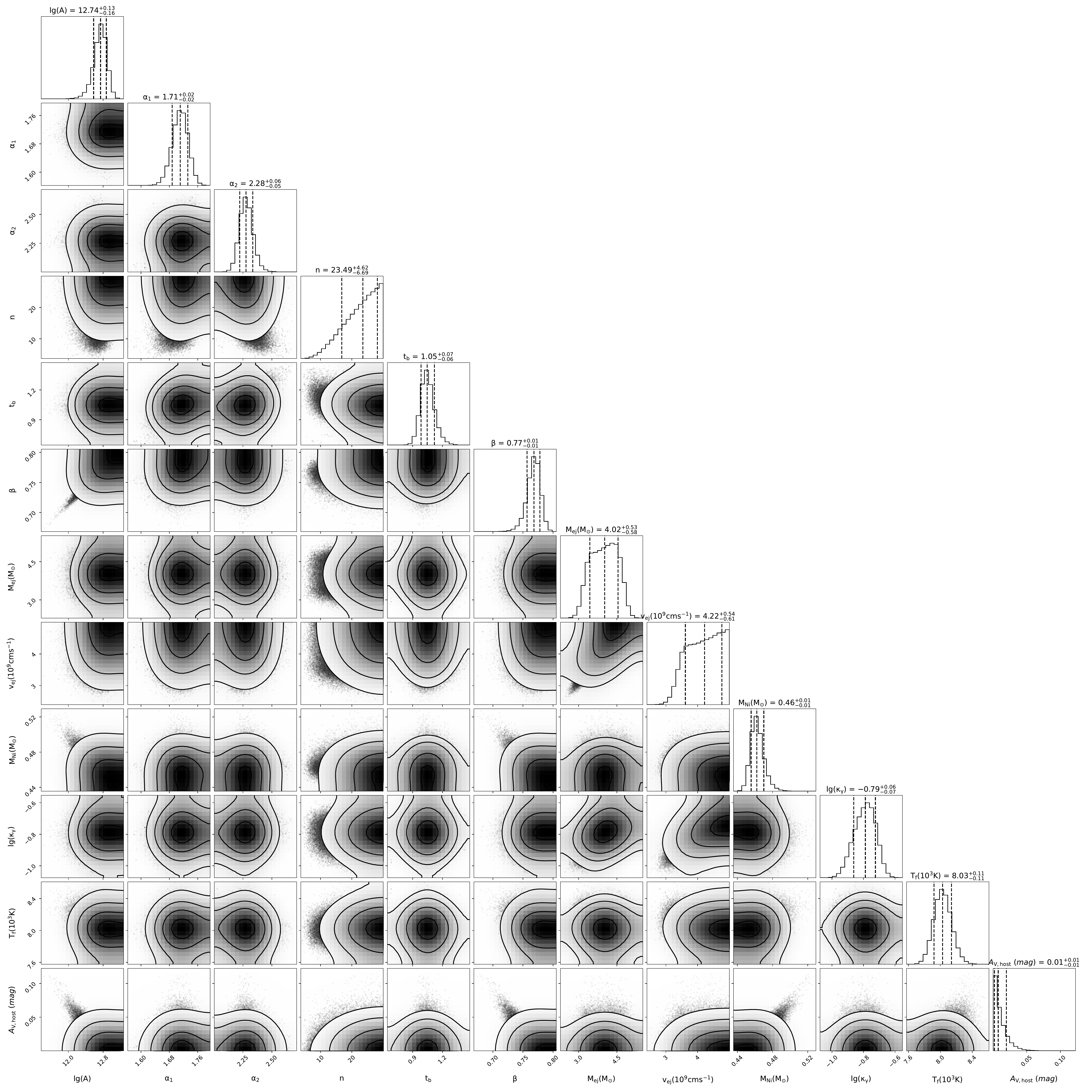}
\end{center}
\caption{The corner plots of the \Ni model for multi-band light curves of GRB~011121/SN~2001ke.}
\label{corner_GRB011121}
\end{figure}

\begin{figure}[tbph]
\begin{center}
\includegraphics[angle=0,width=0.8\textwidth]{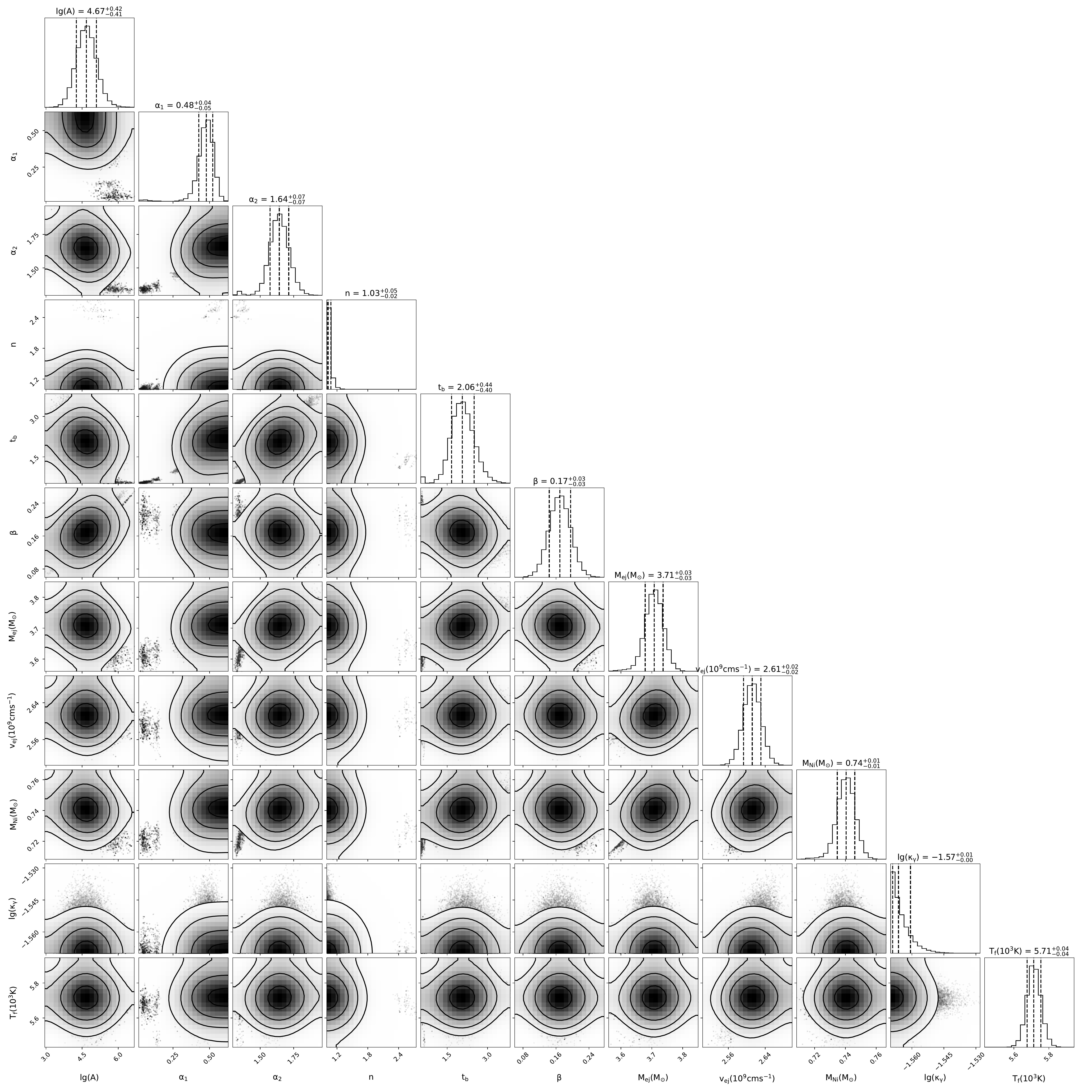}
\end{center}
\caption{The corner plots of the \Ni model for multi-band light curves of GRB~130702A/SN~2013dx.}
\label{corner_GRB130702A}
\end{figure}

\begin{figure}[tbph]
\begin{center}
\includegraphics[angle=0,width=0.8\textwidth]{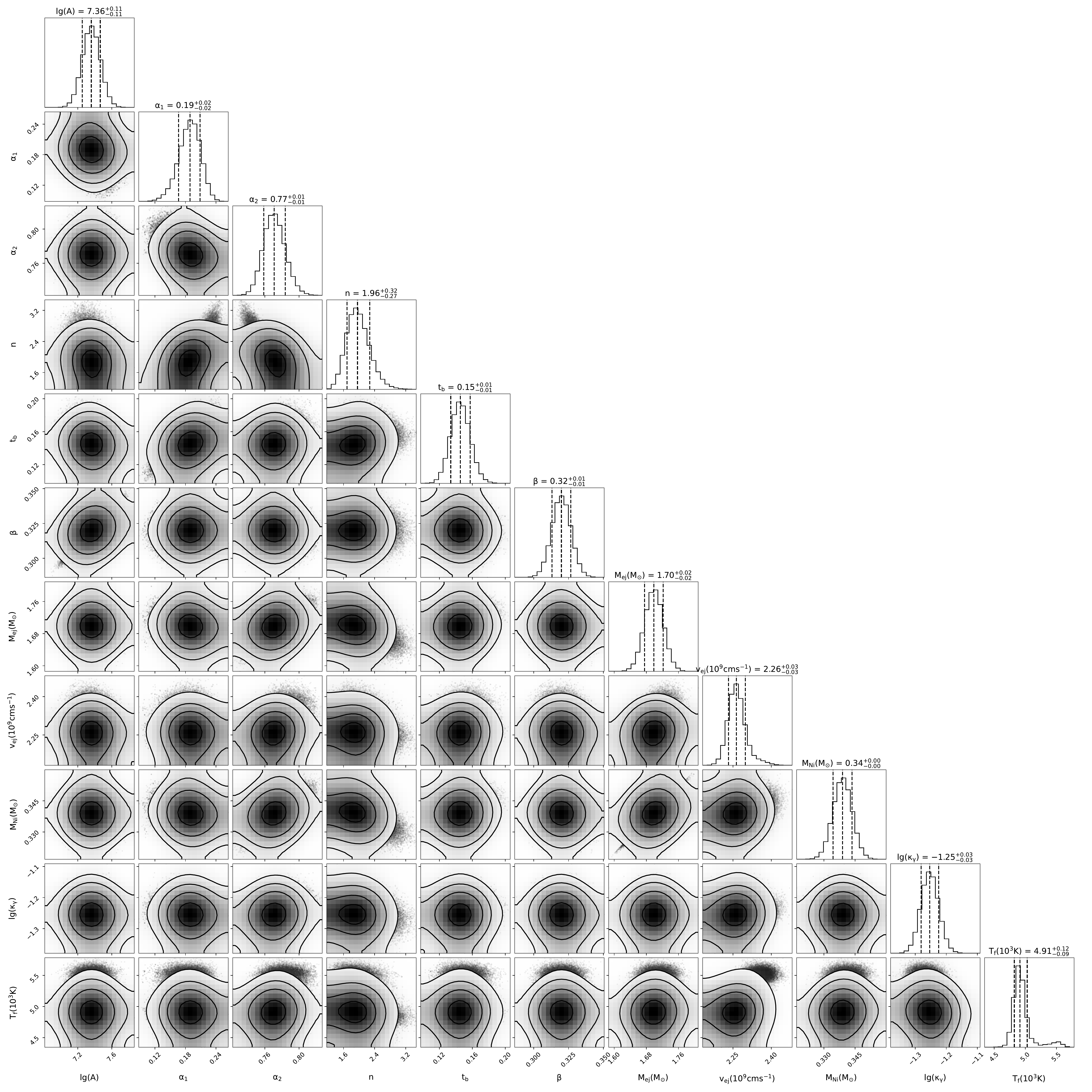}
\end{center}
\caption{The corner plots of the \Ni model for multi-band light curves of GRB~161219B/SN~2016jca.}
\label{corner_GRB161219B}
\end{figure}

\begin{figure}[tbph]
\begin{center}
\includegraphics[angle=0,width=0.8\textwidth]{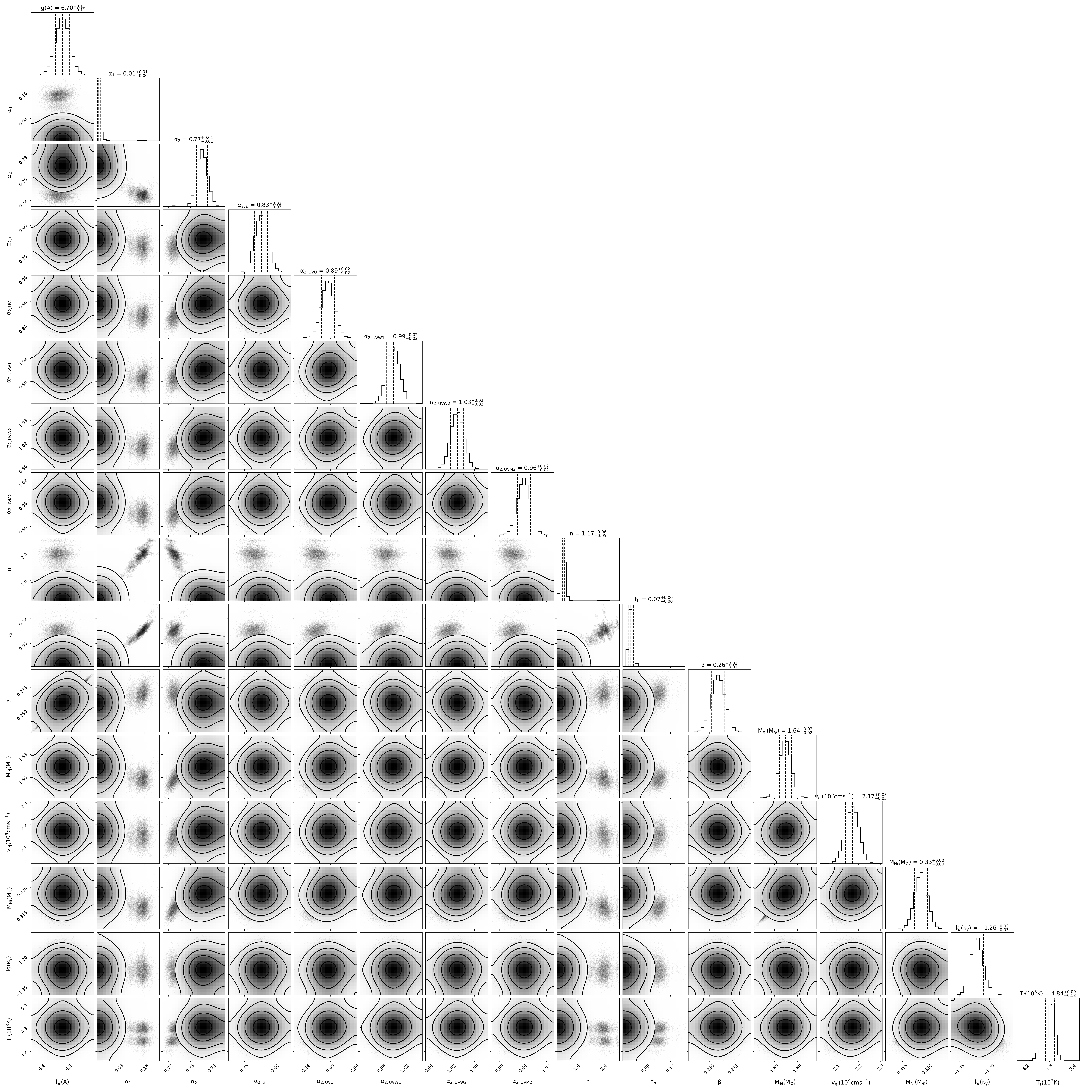}
\end{center}
\caption{{The corner plots of the \Ni model for multi-band light curves of GRB~161219B/SN~2016jca  (assuming
that the $\alpha_2$ values in UV bands are different from each other, and different from that in optical and
NIR bands).}}
\label{corner_GRB161219B-new}
\end{figure}

Figures \ref{corner_GRB011121}, \ref{corner_GRB130702A}, \ref{corner_GRB161219B}, {and \ref{corner_GRB161219B-new}} show the corner plots of the
model for GRB~011121/SN~2001ke, GRB~130702A/SN~2013dx, and GRB~161219B/SN~2016jca  {(two cases)} in the main text.

\clearpage

\clearpage

\clearpage

\end{document}